\documentclass{ws-procs975x65}

\begin{document}

\title{ULTRA-RELATIVISTIC GRAZING COLLISIONS OF BLACK HOLES}

\author{U. SPERHAKE$^{1,*}$, V. CARDOSO$^{2,3}$, F. PRETORIUS$^4$, E. BERTI$^2$, T. HINDERER$^1$\\ and N. YUNES$^4$}

\address{${}^1$Theoretical Astrophysics, California Institute of Technology,
Pasadena, CA 91125, USA\\
${}^2$Department of Physics and Astronomy, The University of Mississippi,\\
University, MS 38677, USA\\
${}^3$CENTRA, Dept. de F\'isica, Inst. Sup. T\'ecnico, Av. Rovisco Pais 1,
Lisbon, 1049, Portugal\\
${}^4$Department of Physics, Princeton University, Princeton, NJ 08544, USA\\
$^*$E-mail: sperhake@tapir.caltech.edu\\}

\begin{abstract}
We study gravitational wave emission, zoom-whirl behavior and
the resulting spin of the remnant black hole in highly boosted collisions
of equal-mass, non spinning black-hole binaries with generic impact parameter.
\end{abstract}


\bodymatter

\section*{}\label{sec: intro}
\noindent {\bf \em Introduction:}
Ultra-relativistic black hole (BH) scattering simulations are of high
interest for a variety of areas in contemporary physics.
These include attempts to resolve the
hierarchy problem by adding extra dimensions \cite{Arkani-Hamed1998,
Antoniadis1998, Randall1999}{},
the resulting possibility to produce BHs in particle colliders
and ultra high-energy cosmic ray interactions with the atmosphere
\cite{Banks1999, Eardley2002}{}, the AdS/CFT correspondence
\cite{Maldacena1997, Witten1998, Gubser1998}
as well as fundamental aspects of black-hole dynamics.
Here we summarize results obtained for highly boosted collisions
of BH binaries with generic impact parameter \cite{Sperhake2009short}{};
see also Refs.~\cite{Sperhake2008short, Shibata2008}{}.

\noindent {\bf \em Simulations:}
We have performed our numerical simulations using the {\sc Lean} code
\cite{Sperhake2006}
which employs the {\em moving puncture} method
\cite{Baker2006short, Campanelli2006}{}. The code
is based on the {\sc Cactus} toolkit \cite{Cactusweb}{},
uses {\sc Carpet} \cite{Schnetter2004, Carpetweb} mesh-refinement,
a spectral solver \cite{Ansorg2004}
for BH initial data and Thornburg's {\sc AHFinderDirect}
\cite{Thornburg1996, Thornburg2004}{}.
For more details see \cite{Sperhake2006}{}.
We set up a coordinate system such that
the BHs start on the $x$-axis separated by a coordinate distance $d$
and with radial (tangential) momentum $P_x$ ($P_y$). The impact parameter
is $b\equiv L/P = P_y d/P$, where $P$ is the linear momentum of either
hole and $L$ the initial angular momentum. Our set of simulations
consists of three sequences, characterized by $d$ and
the Lorentz boost $\gamma \equiv (1-v^2)^{-1/2}$: (1) $\gamma=1.520$
$(v=0.753)$ and $d/M=174.1$; (2) $\gamma=1.520$ $(v=0.753)$ and
$d/M=62.4$; (3) $\gamma=2.933$ $(v=0.940)$ and $d/M=23.1$,
where $M$ is the total BH mass. Along each sequence we increase the impact
parameter starting from the head-on limit $b=0$.

\noindent {\bf \em Results:}
For all sequences we identify three distinct regimes in the
$b$ parameter space: (i) {\em immediate} mergers, (ii) {\em non-prompt}
mergers and (iii) the {\em scattering} regime where no common apparent horizon
forms. These regimes are separated by two special values of $b$,
the {\em threshold of immediate merger} $b^*$ and the {\em scattering
threshold} $b_{\rm scat}$.
Roughly speaking, for $b<b^*$,
the holes merge within the first encounter, whereas for $b^*<b<b_{\rm scat}$
they do not, but radiate enough energy to enter a bound state that {\em
eventually} results in merger.
%
\begin{figure}[b]%
\begin{center}
 \parbox{1.3in}{\epsfig{clip=true,figure=crit_traj.eps,width=1.4in}}
 \hspace*{4pt}
 \parbox{3.3in}{\epsfig{clip=true,figure=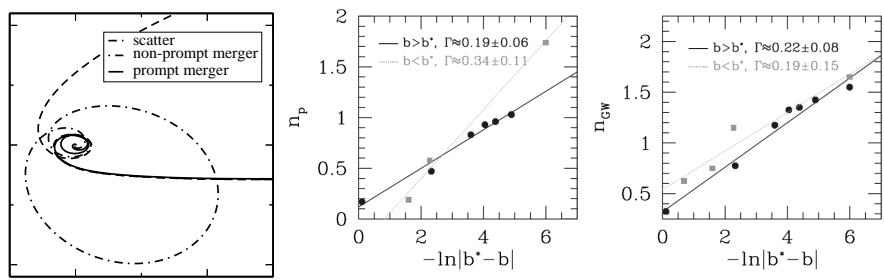,width=3.2in}}
 \caption{Left panel: Puncture trajectories for an immediate merger
          ($b[=3.34~M]<b^*$), a non-prompt merger
          ($b^*<b[=3.39~M]<b_{\rm scat}$) and
          a scattering orbit ($b[=3.40]~M>b_{\rm scat}$).
          Center and right panel: Estimated number of orbits $n_p$
          and $n_{\rm GW}$ as functions of distance from $b^*$ for
          immediate and non-prompt merger cases in sequence 1.
         }
\label{fig1}
\end{center}
\end{figure}
In the left panel of
Fig.~\ref{fig1} we illustrate these regimes by plotting
the trajectories of one binary member, respectively,
for three simulations of sequence 1 with $b/M=3.34$, $3.39$
and $3.40$.
Analysis of the entire sequences 1 and 2 shows that merger occurs only
for $b/M<3.4$, consistent with the estimate $b_{\rm crit}/M \sim
(2.5\pm 0.05)/v$ of Shibata et al.\cite{Shibata2008}{}. For sequence 3, however,
we obtain $2.3 \lesssim b_{\rm scat}/M \lesssim 2.4$,
indicating that Ref.\cite{Shibata2008}
might overestimate $b_{\rm scat}$ for large $\gamma$.
We emphasize in this context the excellent agreement of our findings
with the point-particle approximation\cite{Berti2010}{};
a cross section estimate from
high-energy scattering off a Kerr BH with $j\simeq 0.98$ gives $b_{\rm scat}/M
\simeq 2.36$.

The notion of a {\em threshold of immediate merger} arises in the
context of the geodesic limit.
The argument is that this
threshold should generically be accompanied by behavior akin to
{\em zoom-whirl} orbits in the geodesic limit \cite{Pretorius2007,
Grossman2008, Healy2009a}{}. The number
of ``whirls'' exhibits exponential dependence on the impact parameter
according to $n=C-\Gamma \ln |b-b^*|$, where $C$ is a constant
and $\Gamma$ is inversely proportional to the Lyapunov instability
exponent of the limiting spherical orbit; see Ref.\cite{Pretorius2007}
for details.
We have analyzed sequence 1 by estimating the number of orbits $n$ in
the whirl phase using (a) the puncture trajectories ($n_p$) and
(b) the gravitational wave flux ($n_{\rm GW}$); cf.~Fig.~1 in
\cite{Sperhake2009short}{}.
These two estimates are shown in the center and right panel of
Fig.~\ref{fig1} and indicate that the above mentioned relationship
between $n$ and $b$ is valid with a slope $\Gamma\sim 0.2$ to within
$50~\%$.

The threshold of immediate merger $b^*$ appears to also play a
special role when we consider GW emission and final spin in cases
where merger occurs: both quantities show maxima for $b\sim b^*$.
For $v\sim 0.75$ the radiated energy increases from $\sim 2.2~\%$
of the total energy of the system
for $b=0$ to $\gtrsim 23\%$ near $b^*$. Even for this comparatively
small boost we thus exceed the maximum of $14\pm 4~\%$ reported for
ultra relativistic head-on collisions \cite{Sperhake2008short}{}.
For sequence 3 with
$v\sim 0.94$ and $b\sim b^*$ the radiated energy is $35\pm 5~\%$.
Extrapolation to $v=1$ indicates enormous luminosities $\gtrsim 0.1$
corresponding in physical units to $\sim 3.6 \times 10^{58}~{\rm erg~s}^{-1}$.
Similarly we observe a maximum in the final spin of the post-merger hole
near $b^*$. For sequence 2 and $2.7 \lesssim b/M \lesssim b^*/M$
we obtain a dimensionless spin parameter $j_{\rm fin}>0.9$. For example,
$b=3.04~M$ results in $j_{\rm fin}=0.96 \pm 0.03$, close to extremal and
larger than any values reported so far in the literature \cite{Dain2008,
Washik2008short}{}.

\section*{Acknowledgements}
This work was supported by FCT-Portugal through project
PTDC/FIS/ 64175/2006, NSF grants PHY-0745779, PHY-0601459, PHY-0652995,
PHY-090003 and PHY-0900735, the Alfred P. Sloan Foundation and the Sherman
Fairchild foundation to Caltech. Computations were performed at TeraGrid in
Texas, Magerit in Barcelona, the Woodhen cluster at Princeton University and
HLRB2 Garching.

\bibliographystyle{ws-procs975x65}

\begin{thebibliography}{10}

\bibitem{Arkani-Hamed1998}
N.~Arkani-Hamed, S.~Dimopoulos and G.~R. Dvali, {\em Phys. Lett.} {\bf B429},
  263 (1998).

\bibitem{Antoniadis1998}
I.~Antoniadis, N.~Arkani-Hamed, S.~Dimopoulos and G.~R. Dvali, {\em Phys. Lett.
  B} {\bf 436}, 257 (1998).

\bibitem{Randall1999}
L.~Randall and R.~Sundrum, {\em Phys. Rev. Lett.} {\bf 83}, 3370 (1999).

\bibitem{Banks1999}
T.~Banks and W.~Fischler, {\em Phys. Rev. D} {\bf 66}, p. 044011 (2002).

\bibitem{Eardley2002}
D.~M. Eardley and S.~B. Giddings, {\em Phys. Rev. D} {\bf 66}, p. 044011
  (2002).

\bibitem{Maldacena1997}
J.~M. Maldacena, {\em Adv. Theor. Math. Phys.} {\bf 2}, p. 231 (1997).

\bibitem{Witten1998}
E.~Witten, {\em Adv. Theor. Math. Phys.} {\bf 2}, 253 (1998).

\bibitem{Gubser1998}
S.~S. Gubser, I.~R. Klebanov and A.~M. Polyakov, {\em Phys. Lett. B} {\bf 428},
  105 (1998).

\bibitem{Sperhake2009short}
U.~Sperhake~{\em et al.}, {\em Phys. Rev. Lett.} {\bf 103}, p. 131102 (2009).

\bibitem{Sperhake2008short}
U.~Sperhake~{\em et al.}, {\em Phys. Rev. Lett.} {\bf 101}, p. 161101 (2008).

\bibitem{Shibata2008}
M.~Shibata, H.~Okawa and T.~Yamamoto, {\em Phys. Rev. D} {\bf 78}, p. 101501(R)
  (2008).

\bibitem{Sperhake2006}
U.~Sperhake, {\em Phys. Rev. D} {\bf 76}, p. 104015 (2007).

\bibitem{Baker2006short}
J.~G. Baker~{\em et al.}, {\em Phys. Rev. Lett.} {\bf 96}, p. 111102 (2006).

\bibitem{Campanelli2006}
M.~Campanelli, C.~O. Lousto, P.~Marronetti and Y.~Zlochower, {\em Phys. Rev.
  Lett.} {\bf 96}, p. 111101 (2006).

\bibitem{Cactusweb}
{Cactus Computational Toolkit homepage:} {\tt http://www.cactuscode.org/}.

\bibitem{Schnetter2004}
E.~Schnetter, S.~H. Hawley and I.~Hawke, {\em Class. Quantum Grav.} {\bf 21},
  1465 (2004).

\bibitem{Carpetweb}
{Carpet Code homepage}: {\tt http://www.carpetcode.org/}.

\bibitem{Ansorg2004}
M.~Ansorg, B.~Br{\"u}gmann and W.~Tichy, {\em Phys. Rev. D} {\bf 70}, p. 064011
  (2004).

\bibitem{Thornburg1996}
J.~Thornburg, {\em Phys. Rev. D} {\bf 54}, 4899 (1996).

\bibitem{Thornburg2004}
J.~Thornburg, {\em Class. Quantum Grav.} {\bf 21}, 743(21 January 2004).

\bibitem{Berti2010}
E.~Berti~{\em et al.} (2010), arXiv:1003.0812 [gr-qc].

\bibitem{Pretorius2007}
F.~Pretorius and D.~Khurana, {\em Class. Quantum Grav.} {\bf 24}, S83 (2007).

\bibitem{Grossman2008}
R.~Grossman and J.~Levin, {\em Phys. Rev. D} {\bf 79}, p. 043017 (2008).

\bibitem{Healy2009a}
J.~Healy, P.~Laguna, R.~A. Matzner and D.~M. Shoemaker, {\em Phys. Rev. Lett.}
  {\bf 103}, p. 131101 (2009).

\bibitem{Dain2008}
S.~Dain, C.~O. Lousto and Y.~Zlochower, {\em Phys. Rev. D} {\bf 78}, p. 024039
  (2008).

\bibitem{Washik2008short}
M.~C. Washik~{\em et al.}, {\em Phys. Rev. Lett.} {\bf 101}, p. 061102 (2008).

\end{thebibliography}

\end{document}